\documentclass[journal]{IEEEtran}
\usepackage{amsmath,amsfonts}
\usepackage{algorithm}
\usepackage{array}
\usepackage[caption=false,font=normalsize,labelfont=sf,textfont=sf]{subfig}
\usepackage{textcomp}
\usepackage{stfloats}
\usepackage{url}
\usepackage{verbatim}
\usepackage{graphicx}
\usepackage{cite}
\usepackage{tikz}
\usepackage[thinc]{esdiff}
\usepackage{amssymb}
\usepackage{algpseudocode}
\usepackage{tikz}
\usetikzlibrary{positioning,arrows.meta,calc,fit}
\usepackage{xcolor}
\usepackage{amsmath}
\usetikzlibrary{matrix,calc}
\usetikzlibrary{chains,positioning}
\usepackage{multirow}

\begin{document}

\title{Von Mises–Based Uncertainty Quantification for Closely Spaced Automotive Radar Targets.
}

\author {Vinay Kulkarni, V. V. Reddy\\
\footnotesize \textit {\scriptsize {vinay.kulkarni@iiitb.ac.in, vinod.reddy@iiitb.ac.in}}\\
\footnotesize \textit{ International Institute of Information Technology,
Bengaluru, India} \\
}

\maketitle
\begin{abstract}
This work investigates uncertainty-aware deep learning approaches for direction-of-arrival (DOA) estimation in automotive radar, focusing on probabilistic modeling and downstream integration. A circular-statistics-based von Mises (VM) ensemble (ENS) is compared with an evidential deep learning (EDL) framework based on a normal-inverse-gamma formulation, yielding a Student-$t$ predictive distribution in the Euclidean domain. The ENS framework produces angular predictions parameterized by $(\mu_{\mathrm{ens}}, \kappa_{\mathrm{ens}})$, enabling interpretable uncertainty aligned with directional geometry. Performance is evaluated under in-distribution and multiple out-of-distribution conditions using risk–coverage and ROC/AUROC analyses. Results indicate that ENS achieves lower uncertainty under nominal conditions and exhibits stronger sensitivity to severe perturbations, whereas EDL provides smoother uncertainty variation and slightly improved ranking consistency. Importantly, the ENS representation enables direct probabilistic integration into association modules via closed-form VM likelihoods, facilitating a unified detection–tracking pipeline. These findings highlight a trade-off between geometric consistency and statistical generality in uncertainty-aware DOA estimation.
\end{abstract}


\section{Introduction}
Modern automotive radar processing typically transforms the intermediate-frequency (IF) signal into Range--Doppler (RD) and/or Range--Angle (RA) representations and applies a detection stage---most commonly an Ordered-Statistic Constant False Alarm Rate (OS-CFAR) detector---to produce binary, cell-level target decisions within a single coherent processing interval (CPI)~\cite{kay1993fundamentals,richards2014fundamentals,sun2020mimo}. The resulting detections are expressed as point estimates of range, Doppler, and angle (polar coordinates), converted to Cartesian form, and passed to clustering, data association, and tracking. A central limitation of this conventional pipeline is that the detection and estimation outputs are predominantly point-valued and provide no intrinsic, sample-level measure of estimation uncertainty. As a result, downstream association is implicitly burdened with additional filtering and validation, relying on auxiliary quality indicators (e.g., radar cross section (RCS), signal-to-noise ratio (SNR), and kinematic plausibility) before statistical gating (e.g., Mahalanobis-distance-based validation) can be reliably applied.

This limitation becomes more pronounced in millimeter-wave (mmWave) automotive radar systems, which often operate in the optical scattering regime where extended targets produce spatially distributed, aspect-dependent reflections across multiple range, Doppler, and angle cells~\cite{skolnik1980introduction,wang2016spatial,barton1997radar}. Consequently, a single physical object may give rise to multiple detections within an RD or RA map, while closely spaced objects may generate partially overlapping signatures, particularly in short-range sensing scenarios relevant to (Adavanced Driver Assistance Systems) ADAS~\cite{series2014systems}. In practical systems, ambiguity in multi-target presence is frequently addressed using angle-estimation quality heuristics, such as the Rayleigh resolution criterion or the 3-dB width of angular spectrum peaks, to infer target separability. However, these measures provide only indirect proxies for estimation reliability and can be inadequate under low SNR, target proximity, or overlapping extended-target scattering, where peak broadening or merging does not uniquely correspond to angular uncertainty.

The absence of explicit and calibrated uncertainty at the measurement level propagates this ambiguity to downstream processing and becomes especially critical in centralized perception architectures such as the Autonomous Driving Control Unit (ADCU), where measurements from heterogeneous sensors (e.g., radar, LiDAR, and cameras) are fused to support safety-critical functions~\cite{Qian2025SensorFusionSurvey,Dowling2026CentralizedRadar}. In such fusion pipelines, uncertainty information is essential to appropriately weight sensor contributions, manage cross-modal disagreement, and enable robust decision-making under adverse conditions~\cite{Gawlikowski2023Survey}. At the same time, ADCU constraints on latency and computational complexity limit the ability of downstream association and tracking modules to resolve ambiguity over multiple dwells, further motivating uncertainty-aware radar front-end representations that remain efficient and scalable.

Prior research has sought to improve radar front-end robustness in dense automotive environments through both analytical and learning-based extensions. CFAR variants and adaptive reference-window strategies have been proposed to mitigate heterogeneous clutter, clutter edges, and target-interference effects in RD processing while retaining real-time feasibility~\cite{Kazazi2025RDCFAR,CFARComparison2025IEEEAccess,Wang2023AdaptiveCFAR,Rohling2024KACFAR}. In parallel, tracking-level approaches (e.g., JPDA, MHT, and RFS-based filters) explicitly model association ambiguity in closely spaced scenarios~\cite{BarShalom2009Tracking,Mahler2007PHD}, while recent formulations exploit clustering and message-passing techniques to improve robustness in dense target settings~\cite{Yu2025CloselySpacedAssociation}. Learning-based detectors operating on RD/RA representations and cross-sensor supervision have further demonstrated improved performance over CFAR in challenging scenes involving extended automotive targets~\cite{Delamou2023DLRadar,Roldan2024SeeFurtherThanCFAR,Bauw2026ComplexNNRadar,Kaiser2021CVNNRadar}. Nevertheless, these approaches largely remain detection- or point-estimation-oriented and typically yield binary decisions or point outputs, without providing calibrated and physically interpretable uncertainty—particularly for angular ambiguity arising from closely spaced or spatially extended targets. Related segment-based RD aggregation methods improve detection robustness for extended targets~\cite{kulkarni2025kan,kulkarni2025gamma,wei2022area,kulkarni2023detection}, but RD-only evidence is insufficient to resolve angular proximity, leaving angular ambiguity to be handled implicitly in downstream processing.

Collectively, these methods address detection robustness and association ambiguity at different stages of the processing pipeline; however, they predominantly treat uncertainty implicitly or downstream, rather than modeling it explicitly at the measurement and angular estimation stage within a single CPI. This gap has motivated the development of probabilistic neural frameworks that explicitly attach uncertainty to learning-based estimators.

Bayesian DOA methods and deep unfolding approaches improve robustness by incorporating hierarchical priors and adaptive noise modeling~\cite{Liang2025NoiseIntegralSBL,Li2024BayesianUnfoldedDOA,Chen2020SBLDOA}. Evidential deep learning (EDL) provides a scalable alternative by learning distributional parameters that induce predictive uncertainty and is widely adopted as a baseline in uncertainty-aware deep models~\cite{Sensoy2018EDL,Amini2020EvidentialRegression,Gawlikowski2023Survey}. Practical techniques such as Monte Carlo dropout and deep ensembles are also commonly used for uncertainty benchmarking~\cite{Gal2016MCDropout,lakshminarayanan2017simple}. However, generic uncertainty parameterizations do not explicitly encode the circular statistics inherent to angular variables. This limitation motivates angular uncertainty models that are both probabilistic and structurally consistent with circular statistics, leading naturally to a Von Mises–based formulation that yields physically interpretable dispersion estimates for closely spaced and extended automotive radar targets.

The main contributions of this paper are summarized as follows:
\begin{itemize}
    \item A task-aligned framework for \emph{focused DOA inference} is introduced using localized region-of-interest (ROI) representations, enabling reliable angular estimation for extended and closely spaced automotive radar targets.

    \item Angular measurements are probabilistically modeled using a Von Mises distribution, yielding circular-statistics-consistent DOA estimation with physically interpretable uncertainty through the concentration parameter.

    \item A probabilistic neural network realizing the Von Mises angular model is designed with spatial and channel attention, providing \emph{fine-grained DOA estimation with uncertainty}. Ensembles of such models are interpreted as an approximation to Bayesian Model Averaging (BMA)~\cite{lakshminarayanan2017simple,wilson2020bayesian}.

    \item A product-of-experts formulation is employed at the measurement level to quantify confidence among competing angular hypotheses (e.g., single-target versus closely spaced multi-target cases), naturally aligning uncertainty estimates with downstream association.

    \item The proposed approach is evaluated under both in-distribution and out-of-distribution conditions and benchmarked against evidential deep learning (EDL).
\end{itemize}

\section{Problem Statement}

\begin{figure}
    \centering    
    \includegraphics[trim={1.75cm 0.3cm 2.2cm 0.5cm},width=0.5\linewidth]{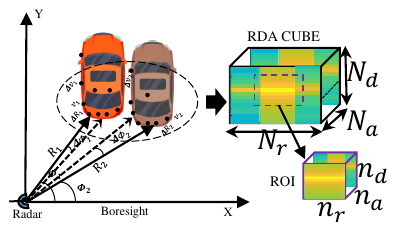}   
    \caption{Closely spaced extended targets and their localized ROI representations in the range--Doppler--angle (RDA) cube within a single CPI.}
    \label{fig:close_proximity}
\end{figure}

In mmwave automotive radar, a physical target is typically composed of multiple distributed scatterers rather than a single point reflector, which makes the target response distributed across neighboring range, Doppler, and angle bins. For the $i^{\text{th}}$ scatterer of the $k^{\text{th}}$ target, the induced perturbations $(\Delta R_{k,i}, \Delta v_{k,i}, \Delta \phi_{k,i})$ produce local deviations around the nominal target state $(R_k, v_k, \phi_k)$. As a result, a single physical object appears as a structured spread in the range--Doppler--angle (RDA) cube rather than as an isolated peak.

In a MIMO radar, the $i^{\text{th}}$ scatterer of the $k^{\text{th}}$ target is located at
\begin{equation}
\phi_{k,i} = \phi_k + \Delta \phi_{k,i},
\end{equation}
where $\phi_k$ denotes the nominal DOA of the target and $\Delta \phi_{k,i}$ captures the scatterer-induced angular deviation. Let
\begin{equation}
\mathbf a(\phi_{k,i})=
\begin{bmatrix}
1,\;
e^{-j\frac{2\pi}{\lambda}d\sin\phi_{k,i}},\;
\ldots,\;
e^{-j\frac{2\pi}{\lambda}(N_a-1)d\sin\phi_{k,i}}
\end{bmatrix}^{T}
\in\mathbb{C}^{N_a}
\label{eq:steering_vector_compact}
\end{equation}
denote the corresponding virtual-array steering vector, where $d$ is the inter-element spacing and $\lambda$ is the wavelength. Then, after beam steering to scan angle $\phi$, the resulting range--Doppler--angle response can be expressed as
\begin{align}
Z(f_R,f_D,\phi)
&=
\sum_{k=1}^{K}\sum_{i=1}^{I_k}
\beta_{k,i}\,
A(\phi;\phi_{k,i})
\delta\!\big(f_R-f_{R_k}-\Delta f_{R_{k,i}}\big)
\nonumber\\
&\quad\times
\delta\!\big(f_D-f_{D_k}-\Delta f_{D_{k,i}}\big)
+\eta(f_R,f_D,\phi),
\label{eq:problem_RDA_cube_final}
\end{align}

where
\begin{equation}
A(\phi;\phi_{k,i}) \triangleq \mathbf a^H(\phi)\mathbf a(\phi_{k,i}),
\label{eq:angular_response_compact}
\end{equation}
denotes the angular response of the $i^{\text{th}}$ scatterer at scan angle $\phi$.
where $\mathbf a(\phi_{k,i})$ denotes the virtual-array steering vector of the $i^{\text{th}}$ scatterer, and $A(\phi,\phi_{k,i})$ represents its beamformed angular response at scan angle $\phi$.

Equations \eqref{eq:problem_RDA_cube_final}--\eqref{eq:angular_response_compact} show that the observed target signature is inherently a localized volume in the RDA cube formed by the superposition of distributed scatterers. As illustrated in Fig.~\ref{fig:close_proximity}, this localized spread may originate from either a single extended target or multiple closely spaced targets with similar kinematic states, making the number of contributing targets itself ambiguous.

Despite the availability of this structured information, conventional processing typically applies CFAR on individual RD/RA cells, yielding sparse detections that do not preserve the underlying scatterer spread. These detections are then filtered using heuristic criteria such as SNR or RCS thresholds before clustering and track association. Such a pipeline presents three key limitations: i) the native target structure in the RDA cube is discarded and only approximately recovered through clustering, ii) a CFAR detection does not indicate whether the localized response corresponds to one or multiple targets, and iii) conventional detections do not provide calibrated measurement-level uncertainty for principled association.

Therefore, the problem is to leverage the localized RDA volume itself, rather than isolated CFAR detections, to infer the underlying target configuration within a single CPI. Specifically, the objective is to determine whether an ROI contains one or more targets, estimate the associated DOA hypothesis (or hypotheses), and quantify the corresponding angular uncertainty in a form suitable for seamless downstream association.

\section{Proposed Approach}
\subsection{Construction of ROI cube}
For automotive vehicles, extended scattering distributes target energy across multiple range--Doppler cells. 
Accordingly, a localized segment of the RD map is extracted to cover approximately \(6\,\mathrm{m}\) in range 
and \(2\,\mathrm{m/s}\) in Doppler, corresponding to \(n_r\times n_d\) bins~\cite{kulkarni2025gamma,kulkarni2025kan}. 
This segment can contain either a single extended vehicle or two closely spaced vehicles with similar kinematics, 
in which case RD-only information is insufficient to reliably determine the number of physical targets.

To overcome this limitation, the extracted RD segment is augmented with the virtual array channels described in~\eqref{eq:steering_vector_compact} and~\eqref{eq:problem_RDA_cube_final}. Incorporating the angle dimension results in a multi-channel measurement structure that exploits the spatial aperture of the virtual array. This extended representation provides additional angular discrimination capability, enabling improved separation between a single extended target and multiple closely spaced targets.

An illustrative example is shown in Fig.~\ref{fig:close_proximity}, where the extracted RD segments embedding the scattering spread of vehicles are stacked across the virtual channels to form a three-dimensional region of interest (ROI) cube. This ROI extracted from RDA cube captures the joint range–Doppler–angle scattering characteristics associated with either a single extended target or multiple closely spaced targets within the scene.

Accordingly, the resulting ROI cube
$\mathcal{Z}_{\mathrm{ROI}} \in \mathbb{C}^{n_d \times n_r \times n_a}$  with pdf $f(z)$ is given by
\begin{align}
\mathcal{Z}_{\mathrm{ROI}}[\ell,n,m]
&=
\sum_{k=1}^{K}
A_m(\phi_k)\,
Y_k[\ell,n], \nonumber \\
\qquad
(\ell,n,m) &\in [1\!:\!n_d] \times [1\!:\!n_r] \times [1\!:\!n_a],
\label{eq:ROI_patch}
\end{align}
where $n_a = N_a$ is chosen to retain the full set of direction-of-arrival (DOA)
channels and thereby preserve complete angular information.
The ROI cube may contain scattering contributions from
$K \in \{0,1,2\}$ physical targets within the selected region, leading to three
mutually exclusive hypotheses corresponding to the number of targets present:
\[
H_0:\,K=0, \qquad
H_1:\,K=1, \qquad
H_2:\,K=2.
\]
\subsection{Product of Experts (PoE)}
The discriminative structure across the $n_a$ virtual array channels encodes directional information relevant for estimating the target’s direction-of-arrival (DOA) from the region of interest $\mathcal{Z}_{\mathrm{ROI}}$. 
To represent azimuth angle estimates together with their associated uncertainty, the von Mises (VM) distribution~\cite{mardia2000directional}, a circular probability distribution on the unit circle, is employed. 
The azimuth angle $\phi$ is restricted to the field of view $\{-90^\circ, +90^\circ\}$ and is modeled as
\begin{equation}
    \mathrm{VM}(\phi \mid \mu,\kappa)
    =
    \frac{\exp\!\big(\kappa \cos(\phi - \mu)\big)}
         {2\pi\, I_0(\kappa)},
    \label{eq:VM_defn}
\end{equation}
where $\mu$ denotes the mean direction, $\kappa \ge 0$ is the concentration parameter controlling angular dispersion, and $I_0(\kappa)$ is the modified Bessel function of the first kind and order zero.

As the region-of-interest (ROI) cube $\mathcal{Z}_{\mathrm{ROI}}$ is constructed to
contain scattering contributions from at most two physical targets, the angular
density is represented using a maximum of two von Mises (VM) components.
Accordingly, a Product-of-Experts (PoE) formulation~\cite{hinton2002training} is
adopted, in which the individual angular components are assumed to be
independent and identically distributed (i.i.d.), and the joint density is given
by the normalized product of the VM experts.

Target absence or presence is modeled through the concentration parameters of the
VM components. Setting $\kappa=0$ yields a uniform angular distribution, indicating
complete uncertainty in direction and effectively disabling the corresponding
expert. This mechanism enables a unified representation of no-target ($K=0$),
single-target ($K=1$), and two-target ($K=2$) cases without altering model
structure.

Under this formulation, the joint angular densities corresponding to the three
hypotheses are
\begin{align}
f\big(z=\mathcal{Z}_{\mathrm{ROI}}\big)
=
\begin{cases}
\text{H0}:
\mathrm{VM}_1(\phi \mid \mu_1, 0)\;
\mathrm{VM}_2(\phi \mid \mu_2, 0), \\[5pt]
\text{H1}:
\mathrm{VM}_1(\phi \mid \mu_1, \kappa_1)\;
\mathrm{VM}_2(\phi \mid \mu_2, 0), \\[5pt]
\text{H2}:
\mathrm{VM}_1(\phi \mid \mu_1, \kappa_1)\;
\mathrm{VM}_2(\phi \mid \mu_2, \kappa_2),
\end{cases}
\label{eq:poe_vm}
\end{align}
where $\mu_1,\mu_2$ denote the DOAs of the potential targets and
$\kappa_1,\kappa_2$ are their corresponding concentration parameters.

Direct multiplication of sharply peaked VM densities can result in numerical
instabilities during training, including gradient explosion for large
$\kappa_i$ and gradient vanishing for small $\kappa_i$. To alleviate these issues
and allow explicit expert activation control, a log-domain PoE formulation is
used:
\begin{align}
\log f(\phi \mid z)
&\propto \sum_{i=1}^{2} w_i(z)\,
\log \mathrm{VM}_i\!\left(\phi\mid\mu_i(z),\kappa_i(z)\right),
\label{eq:log_poe_vm}
\end{align}
where $w_i(z)\in[0,1]$ are data-dependent pooling weights that modulate the
contribution of each expert.

\subsection{Bayesian approximation of DOA Estimation}
\label{sec:bayes_DOA}

Given an ROI cube \(z \equiv \mathcal{Z}_{\mathrm{ROI}}\), the objective is to estimate the posterior DOA density \(f(\phi \mid z)\). 
Under a non-informative (uniform) prior over the field of view ($-90^\circ \text{ to } 90^\circ$), the posterior density approximates to the likelihood:
\begin{align}
f(\phi \mid z)
\;\propto\;
\frac{1}{\pi} f(z \mid \phi) 
\label{eq:posterior_likelihood_equiv}
\end{align}
and DOA learning is implemented by minimizing the negative log-likelihood of the Von--Mises (PoE) model (Section~\ref{sec:loss}).

To further capture model uncertainty, an ensemble of VM-based probabilistic models (ENS) is employed. 
The aggregated predictive distribution over \(\phi\) can be interpreted as an approximation to Bayesian model averaging (BMA), 
where different ensemble members act as model realizations and the final prediction is obtained by combining their respective likelihoods. 
This enables the representation of both aleatoric uncertainty (through the Von--Mises model) and epistemic uncertainty arising from variability across model realizations.
\subsection{Probabilistic Neural Network Design}
\begin{figure}[htbp]
    \centering
    \includegraphics[trim={2.5cm 0.2cm 3.3cm 0.5cm},width=0.40\linewidth,height=0.38\linewidth]{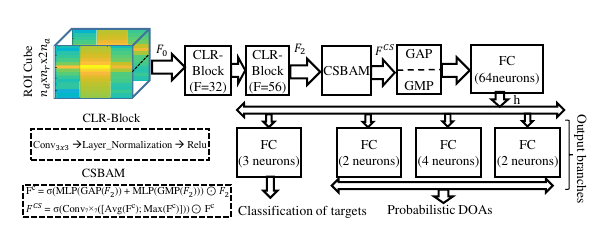}
    \caption{Architecture of the proposed probabilistic neural network.
    The network branches into (i) a target classification head producing target presence probabilities
    $\bar{P_{r,k}}(z)$, and (ii) a probabilistic DOA head with von Mises experts, estimating pooling weights
    $\{w_i\}$,concentration parameters $\{\kappa_i\}$, and azimuth directions
    $\{\mu_i\}_{i=1}^{2}$. The azimuth $\mu_i$ is parameterized via Cartesian
    components $(u_{x_i},v_{y_i})$, with
    $\mu_i = \tan^{-1}(v_{y_i}/u_{x_i})$.
    }
    \label{fig:BNN}
\end{figure}

After defining the PoE model and the target-count framework, the input to the network (Fig.~\ref{fig:BNN}) is the ROI cube \( \mathcal{Z}_{\mathrm{ROI}} \).
To account for variations in signal strength due to target distance and SNR, an energy-based normalization is applied:
\begin{align}
\mathrm{rms} &=
\sqrt{\frac{1}{n_r n_d n_a} \sum_{j=1}^{n_r n_d n_a} |z_j|^2}, \\
\mathcal{Z}_{\mathrm{ROI}} &\leftarrow \frac{\mathcal{Z}_{\mathrm{ROI}}}{\mathrm{rms}}.
\end{align}

The real and imaginary components are concatenated to form the input tensor 
\(F_0 \in \mathbb{R}^{n_d \times n_r \times 2n_a}\).
Feature extraction is performed using two convolutional blocks (Conv$_{\text{kernel size}}^{\text{(Filters)}}$-LayerNorm(LN)–ReLU), followed by a channel–spatial attention module:
\begin{align}
F_2 = \text{ReLU}\!\left(\text{LN}\!\left(\text{Conv}^{(56)}_{3\times3}
\left[\text{ReLU}\!\left(\text{LN}\!\left(\text{Conv}^{(32)}_{3\times3}(F_0)\right)\right)\right]\right)\right).
\end{align}

\subsubsection{Channel and Spatial Attention}

To enhance discriminative features for closely spaced targets, a lightweight
channel–spatial attention mechanism based on CBAM~\cite{woo2018cbam} is
incorporated. Channel attention adaptively reweights feature channels using
global average pooling (GAP) and global max pooling (GMP), while spatial
attention emphasizes the salient range--Doppler regions.

The combined attention operation is defined as
\begin{align}
F^{C} &=
\sigma\!\Big(
\text{FC}^{(2)}(\text{GAP}(F_2)) +
\text{FC}^{(16)}(\text{GMP}(F_2))
\Big)\odot F_2, \\
F^{CS} &=
\sigma\!\Big(
\text{Conv}_{7\times7}
([\text{Avg}(F^{C});\text{Max}(F^{C})])
\Big)\odot F^{C}.
\end{align}

Here, $\text{FC}^{(n)}(\cdot)$ denotes a fully connected layer with $n$ neurons.
$\text{Avg}(\cdot)$ and $\text{Max}(\cdot)$ denote average and max pooling along
the channel dimension, respectively. The channel-attended feature map $F^{C}$ is
subsequently used as input to the spatial attention branch.

These refined features $F^{CS}$ are aggregated via global pooling to obtain a
compact latent representation,
\begin{align}
h =
\text{ReLU}\!\big(
\text{FC}^{(64)}([\text{GAP}(F^{CS});\text{GMP}(F^{CS})])
\big).
\end{align}

\subsubsection{Output Branches}

As illustrated in Fig.~\ref{fig:BNN}, the network comprises two output branches.

\noindent\textit{(i) Classification head:}
Given the hypothesis-dependent scores produced by the classification head, we interpret these as proportional to the likelihood terms \(f(z \mid H_k)\). 
Assuming uniform priors over \(\{H_0,H_1,H_2\}\), the normalized target-presence probabilities are defined as
\begin{align}
\bar{P}_{r,k}(z)
&\triangleq
\frac{ f(z \mid H_k) }
     { \sum_{j=0}^{2} f(z \mid H_j) }, \quad k \in \{0,1,2\}, \nonumber \\
&= \big[\mathrm{softmax}(\mathrm{FC}^{(3)}(h))\big]_k .
\label{eq:Pr_likelihood_norm}
\end{align}

Here, \(f(z \mid H_k)\) denotes a learned discriminative score proportional to the likelihood of observing \(z\) under hypothesis \(H_k\). The softmax operation performs normalization and yields posterior probabilities under an implicit uniform prior.
The estimated number of targets is
\begin{equation}
\widehat{K}
=
\arg\max_{k\in\{0,1,2\}} \bar{P}_{r,k}(z).
\end{equation}

The classification branch is trained using cross-entropy loss, 
which corresponds to the negative log-likelihood of the categorical distribution over the target-presence hypotheses. 
In parallel, the probabilistic DoA branch models the conditional angular density \(f(\phi \mid z, H_k)\) and is trained via the negative log-likelihood (NLL) of the Von--Mises Product-of-Experts model to capture angular uncertainty.

\noindent\textit{(ii) Probabilistic DOA head:}
Direction-of-arrival (DOA) estimation and associated uncertainty are modeled
using the log-domain PoE formulation in~\eqref{eq:log_poe_vm},
parameterized by $\{w_i,\mu_i,\kappa_i\}_{i=1}^{2}$.
The expert pooling weights and concentration parameters are estimated as
\begin{align}
[w_1,w_2]^{\mathsf{T}} &= \mathrm{softmax}(\mathrm{FC}^{(2)}(h)), \\
[\kappa_1,\kappa_2]^{\mathsf{T}} &= \mathrm{FC}^{(2)}(h).
\end{align}

To account for angular periodicity, azimuth directions are represented in
Cartesian form and subsequently mapped to polar coordinates:
\begin{align}
[u_{x_1},u_{x_2},v_{y_1},v_{y_2}]^{\mathsf{T}} &= \mathrm{FC}^{(4)}(h), \\
\mu_i &= \tan^{-1}\!\left(\frac{v_{y_i}}{u_{x_i}}\right), \quad i=1,2 .
\end{align}

During inference, an ensemble of independently trained von Mises models is
employed to enable Bayesian model averaging (BMA), with epistemic uncertainty
captured through variability across ensemble members.

\subsection{Loss Function}
\label{sec:loss}

The proposed network is trained to jointly optimize target-presence classification 
and conditional DOA estimation. For a mini-batch of size \(B\), the total training 
objective is defined as
\begin{align}
\mathcal{L}
&=
\frac{\lambda_{\mathrm{CE}}}{B}\sum_{i=1}^{B}
\mathcal{L}_{\mathrm{CE_i}}
+
\frac{\lambda_{\mathrm{NLL}}}{|S|}\sum_{i\in S}\mathcal{L}_{\mathrm{NLL},i}
\nonumber \\
&\quad +
\mathcal{L}_{\mathrm{couple}}
+
\lambda_{\kappa}\,\mathbb{E}\!\left[
\max\!\left(\kappa_{i,j}-\kappa_{\mathrm{thr}},\,0\right)^2
\right],
\label{eq:total_loss}
\end{align}
where
\[
S = \{\, i : K_i \in \{1,2\} \,\}
\]
denotes the ROI cubes containing at least one physical target in the batch.
When \(|S|=0\), the NLL term is omitted.
The final term imposes a concentration regularization penalty to prevent 
overconfident concentration estimates by penalizing large values of 
\(\kappa_{i,j}\).

\subsubsection{Cross-Entropy (CE) Loss}
\label{sec:ce_loss}

The classification branch outputs softmax probabilities 
\((\bar{P}_{r,0}, \bar{P}_{r,1}, \bar{P}_{r,2})\), as defined in~\eqref{eq:Pr_likelihood_norm}. 
Under the likelihood-based interpretation, these correspond to normalized hypothesis likelihoods. 
The cross-entropy loss is
\begin{equation}
\mathcal{L}_{\mathrm{CE_i}}
=
-\sum_{k=0}^{2}
Pr_{gt,k}\,\log\!\left(\bar{P}_{r,k}\right).
\label{eq:CE_loss}
\end{equation}

\subsubsection{von--Mises Negative Log-Likelihood (NLL)}
\label{sec:nll_vm}

Prior to defining the negative log-likelihood, the concentration parameters 
\(\kappa_j\) predicted by the network are transformed to obtain stable and 
physically meaningful uncertainty estimates:
\begin{align}
\kappa_{\mathrm{eff},j}
&=
\operatorname{clip}\!\Big(
\operatorname{softplus}(\kappa_{j})+1,\,
\kappa_{\min},\,
\kappa_{\max}
\Big)
\nonumber \\
&\times
\frac{\operatorname{clip}(\mathrm{sig\_rms}_i,\,0.5,\,2.0)}{\tau},
\label{eq:kappa_eff}
\end{align}
where \(j \in \{1,2\}\) denotes the two VM experts.

Conditioned on the ROI cube $z \equiv \mathcal{Z}_{\mathrm{ROI}}$, the network
models a conditional angular density $f(\phi \mid z)$ using the log-domain
Product-of-Experts (PoE) formulation in~\eqref{eq:log_poe_vm}:
{\small \begin{align}\label{eq:logpk_vm}
\log f(\phi \mid z)
\propto
\sum_{j=1}^{2}
w_j \Big[
\kappa_{\mathrm{eff},j}\cos(\phi - \mu_j)
-\log(2\pi)
-\log I_0(\kappa_{\mathrm{eff},j})
\Big].
\end{align}
}

For a given ROI cube, up to two ground-truth azimuth angles
\(
\boldsymbol{\phi} = \{\phi_{\mathrm{gt},1}, \phi_{\mathrm{gt},2}\},
\)
with \(K \in \{0,1,2\}\), may be present.
The network predicts two von Mises components
\(\{\mathrm{VM}(\mu_j,\kappa_j)\}_{j=1}^{2}\).

To associate predicted components with ground-truth angles, the
component--to--target log-likelihood matrix
\(\mathbf{L}\in\mathbb{R}^{2\times 2}\) is defined as
\begin{equation}
\mathbf{L}_{t,j}
=
w_j \,\log f(\phi_{\mathrm{gt},t} \mid z;\mu_j,\kappa_j),
\quad
t,j\in\{1,2\}.
\end{equation}

Since the ordering of predicted components is arbitrary, permutation
invariance is enforced via one-to-one assignment. The optimal assignment
maximizes the total log-likelihood and, for the \(2\times2\) case, with two targets, reduces to
\begin{equation}
\max\!\left(
\mathbf{L}_{1,1}+\mathbf{L}_{2,2},\;
\mathbf{L}_{1,2}+\mathbf{L}_{2,1}
\right).
\end{equation}
This maximization is equivalent to the Hungarian assignment
solution~\cite{kuhn1955hungarian}.

Accordingly, the per-sample negative log-likelihood (NLL) is defined as
\begin{equation}
\mathcal{L}_{\mathrm{NLL},i}=
\begin{cases}
-\max\!\left(
\mathbf{L}_{1,1}+\mathbf{L}_{2,2},\;
\mathbf{L}_{1,2}+\mathbf{L}_{2,1}
\right), & K_i = 2,\\[6pt]
-\max_{j\in\{1,2\}}\mathbf{L}_{1,j}, & K_i = 1,\\[6pt]
0, & K_i = 0.
\end{cases}
\end{equation}

\subsubsection{Coupling Regularization}
\label{sec:coupling}

For target-present scenarios with \(K<2\), only a subset of the predicted
VM components should remain active. To suppress spurious activation of
unused components in a sample $i$, a coupling regularizer is introduced:

\begin{align}
\mathcal{L}_{\mathrm{couple}}
&=
\alpha_{0,w}\,\mathbb{E}_{K=0}\![w_{i,1}+w_{i,2}]
+
\alpha_{0,\kappa}\,\mathbb{E}_{K=0}\![\kappa_{i,1}+\kappa_{i,2}]
\nonumber \\
&\quad +
\alpha_{1,w}\,\mathbb{E}_{K=1}\![w_{i,\mathrm{unused}}]
+
\alpha_{1,\kappa}\,\mathbb{E}_{K=1}\![\kappa_{i,\mathrm{unused}}].
\label{eq:lcouple}
\end{align}

where ``unused'' denotes the VM component whose mean direction is not
assigned to any ground-truth DOA under the optimal assignment.
The expectation $\mathbb{E}_{K=k}[\cdot]$ is taken over samples satisfying 
$K_i = k$, for $k \in \{0,1\}$, enforcing suppression of one or both 
components when fewer than two targets are present.
\subsection{Comparison with Uncertainty Estimation Methods}

To evaluate the proposed Von Mises-based probabilistic neural network (PNN), 
two widely used uncertainty estimation approaches are considered: Deep 
Ensembles (ENS) which approximates BMA ~\cite{lakshminarayanan2017simple} and Evidential Deep Learning (EDL).

Deep Ensembles are constructed by training multiple independent 
instances of the complete PNN model (Fig.~\ref{fig:BNN}) with different random 
initializations, and aggregating their predictions and uncertainties using 
circular statistics. This captures epistemic uncertainty through model 
diversity, while inheriting aleatoric uncertainty from the base PNN formulation.

Evidential Deep Learning, on the other hand, retains the same feature 
extraction backbone as the PNN architecture (Fig.~\ref{fig:BNN}), but replaces 
its output branches. Specifically, the classification head is replaced with an 
evidential classifier based on Dirichlet distributions, while the probabilistic DOA regression 
head is replaced with a Normal–Inverse–Gamma (NIG) formulation to model predictive 
uncertainty \cite{Sensoy2018EDL,Amini2020EvidentialRegression,Gawlikowski2023Survey}. 
The network is trained by minimizing the negative log-likelihood of the resulting 
Student-t predictive distribution, along with evidence regularization 
terms, enabling estimation of both epistemic and aleatoric uncertainty. The implementation follows the formulation and public reference code of Deep Evidential Regression \cite{Amini2020EvidentialRegression,amini_edl_github}.
The feature extraction backbone remains unchanged to ensure a fair comparison.

This design enables a comparison of uncertainty estimation 
strategies under a common backbone, isolating the impact of the uncertainty 
modeling approach.
\section{Data Synthesis}

The statistical relationships governing target ranges, velocities, angles, 
and scatterer distributions are described below. The generative process 
for each target \(k\), with \(I_k = 50\) scatterers, is defined as
\begin{align*}
R_k &\sim U(10,75), \qquad v_k \sim U(-19.6,19.6), \\
\phi_k &\sim U(-85^\circ,85^\circ), \\
\phi_{k,2} &= \text{clip}\!\big(\phi_k + U(5^\circ,20^\circ),\, -90^\circ,\,90^\circ\big), \\
R_{k,i} &\sim R_k + U(-1.6,1.6), \\
v_{k,i} &\sim v_k + \mathcal{N}(0,0.3), \\
\phi_{k,i} &\sim 
\begin{cases}
\mathcal{N}(\phi_k,\,5^\circ), & R_k \le 40, \\[4pt]
\mathcal{N}(\phi_k,\,2^\circ), & R_k > 40,
\end{cases} \\
\beta_{k,i}^2 &\sim \chi^2_4.
\end{align*}

Table~\ref{tab:spec} summarizes the FMCW radar parameters, ROI cube settings, and the target–scatterer configuration used for generating closely spaced vehicle scenarios.
\begin{table}[htbp]
\centering
\renewcommand{\arraystretch}{1.15}
\caption{FMCW Radar and Target Synthesis Parameters}
\label{tab:spec}
\begin{tabular}{|p{3.1cm}|p{2.3cm}|p{2.2cm}|}
\hline
\textbf{Parameter} & \textbf{Value} & \textbf{Notes} \\ \hline

$f_0$ & $77$ GHz & Carrier frequency \\ \hline
Chirp slope $\mu$ & $16.67$ MHz/$\mu$s & \\ \hline
$T_{\text{cri}}$ & $50~\mu$s & Chirp duration \\ \hline
$f_s$ & $10$ MHz & Sampling rate \\ \hline
Chirp type & Up-chirp & FMCW \\ \hline
Antenna configuration & ULA & $\lambda/2$ spacing \\ \hline
Direction & Azimuth & \\ \hline
Fast-time samples $(N_a)$ & $256$ & Range bins \\ \hline
Chirps $(N_d)$ & $128$ & Doppler bins \\ \hline
Virtual channels & $16$ & MIMO \\ \hline

\multicolumn{3}{|c|}{\textbf{ROI cube Specifications}} \\ \hline
Range resolution & $0.3516$ m & \\ \hline
Velocity resolution & $0.3044$ m/s & \\ \hline
$\mathcal{Z}_{ROI} \in \mathbb{C}^{n_d \times n_r \times n_a}$ & $\mathbb{C}^{7 \times 17 \times 16}$ & $2.13$ m/s $\times$ $5.98$  m $\times 16 \text{channels}$ \\ \hline

\multicolumn{3}{|c|}{\textbf{Target Synthesis Parameters}} \\ \hline
Scatterer power & $\chi^2_4$ & Swerling-3 \\ \hline
Scatterers per target & $50$ & \\ \hline
Angle difference (2 targets) & $U(5^\circ,20^\circ)$ & Typical lane spacing \\ \hline
Scatterers: Range & $R_k + U(\pm1.6~\text{m})$ & $\sim 3.2$ m width \\ \hline
Scatterers: Velocity & $v_k + \mathcal{N}(0,0.3~\text{m/s})$ & $\approx 1$ Doppler bin \\ \hline
Scatterers: Angle & $\mathcal{N}(\phi_{gt},5^\circ)$ & $R\ge40$: $2^\circ$ \\ \hline
SNR range & $-25$ to $25$ dB & Scene-level (before ROI extraction and scaling) \\ \hline
Noise type & AWGN & \\ \hline
RCS (side view) & $19$–$22$ dBm$^2$ & \\ \hline
RCS (front view) & $8.7$–$20.5$ dBm$^2$ & \\ \hline
RCS (rear view) & $14.4$–$24.6$ dBm$^2$ & \\ \hline
Target distance & $10$–$75$ m & MRR \\ \hline
Target angles & $\phi_{gt}\in[-85^\circ,85^\circ]$ & Side-looking radar \\ \hline
Simulated data(in m) & $25$,$45$ & Radar simulated\\ \hline
Augmented data(in m) & $10,17,33$& \\& $53,60,70$ & Augmented \\ \hline
\end{tabular}
\end{table}

\textbf{Data Augmentation.}
To increase data diversity and improve generalization, each ROI cube of size 
\(7 \times 17 \times 16\), as defined by the radar specifications in 
Table~\ref{tab:spec}, is augmented using range-dependent scaling and 
controlled noise injection. Simulated data generated at reference distances 
of 25\,m and 40\,m serve as base templates for augmentation to other ranges.

Noise is first suppressed using the median magnitude of the ROI cube. 
Range scaling is then applied based on the radar range equation (RRE), 
where the received power scales as \(\beta_k^2 \propto 1/R_k^4\), 
to emulate amplitude variation with distance. 
A small Gaussian smoothing kernel,
\[
\begin{bmatrix}
0.05 & 0.10 & 0.05 \\
0.10 & 0.40 & 0.10 \\
0.05 & 0.10 & 0.05
\end{bmatrix},
\]
is convolved across the range–Doppler plane for all channels to introduce 
controlled spatial spreading across neighboring bins, approximating the sidelobe 
leakage effects arising from finite-resolution FFT processing.

This augmentation pipeline expands the dataset while maintaining statistical 
fidelity to real radar measurements.

\section{Results and Analysis}
The ensemble model (ENS), obtained by aggregating three VM-based probabilistic neural networks trained with different random seeds, achieves a classification accuracy comparable to the EDL baseline (Table~\ref{tab:vm_edl_confusion}). 
While overall accuracies are similar, the VM-based models provide improved class-wise performance for closely spaced targets, with diagonal entries of approximately $90\%$ compared to $87\%$ for EDL, consistently across random initializations.

\begin{table}[t]
\centering
\caption{Performance of VM models (different random seeds) and the EDL model with confusion matrices.
Rows and columns denote ground-truth and predicted classes (0,1,2), respectively.}
\label{tab:vm_edl_confusion}
\small
\setlength{\tabcolsep}{3pt}
\renewcommand{\arraystretch}{1.0}

\begin{tabular}{|c|c|c|}
\hline
\textbf{Model} & \textbf{Acc. (\%)} & \textbf{Confusion Matrix (\%)} \\ \hline

VM (Seed 42) & 92.76 &
$\left[\begin{matrix}
97 & 1 & 2\\
5 & 91 & 4\\
5 & 5 & 90
\end{matrix}\right]$ \\ \hline

VM (Seed 85) & 93.09 &
$\left[\begin{matrix}
98 & 0 & 2\\
5 & 91 & 5\\
5 & 5 & 90
\end{matrix}\right]$ \\ \hline

VM (Seed 12) & 92.04 &
$\left[\begin{matrix}
98 & 1 & 2\\
5 & 91 & 4\\
5 & 7 & 88
\end{matrix}\right]$ \\ \hline

EDL & 91.01 &
$\left[\begin{matrix}
95 & 2 & 3\\
3 & 91 & 6\\
4 & 9 & 87
\end{matrix}\right]$ \\ \hline

\end{tabular}
\vspace{2pt}
\end{table}

Beyond accuracy, reliable uncertainty estimation requires probabilistic characterization. 
Accordingly, ensemble predictions are aggregated using circular statistics from directional statistics \cite{mardia_jupp} to compute epistemic and aleatoric uncertainty measures for DOA estimation. 
In the proposed VM-based probabilistic neural network, epistemic uncertainty reflects variability across ensemble members due to data and initialization effects, whereas aleatoric uncertainty arises from the intrinsic dispersion of the predicted von Mises distributions within each model. 
The complete formulations are provided in Appendix~A. In the following sections, the proposed ENS and the EDL baseline are compared under in-distribution and out-of-distribution conditions.

\subsection{In-Distribution (ID) Analysis}
\begin{figure*}
    \centering
    \includegraphics[trim={1cm 0cm 0cm 0 cm},width=\linewidth, height=0.33\linewidth]{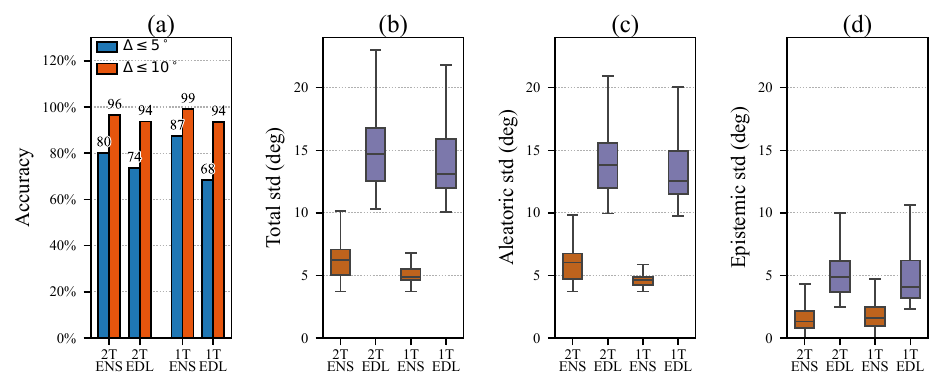}
    \caption{In-distribution performance evaluation using $8602$ closely spaced target samples.
    (a) DOA estimation accuracy of VM-based Deep Ensembles (ENS, 3 models) and Evidential Deep Learning (EDL).
    (b) Total, (c) aleatoric, and (d) epistemic standard deviation under in-distribution conditions.}
    \label{fig:ID_Uncertainty}
\end{figure*}

The in-distribution test samples, which are disjoint from the training set and follow the same statistical characteristics, including the range of target distances specified in Table~\ref{tab:spec}, are considered for evaluation. Representative examples illustrating target presence detection, DOA estimation, and the associated uncertainty measures are summarized in Tables~\ref{tab:id_presence_doa} and~\ref{tab:id_uncertainty}. The overall estimation accuracy and uncertainty behavior under in-distribution conditions are further analyzed in Fig.~\ref{fig:ID_Uncertainty}.

\begin{table}[!t]
\centering
\caption{Representative ID examples: Presence and DOA estimates (ENS vs EDL).}
\label{tab:id_presence_doa}
\footnotesize
\setlength{\tabcolsep}{3pt}
\renewcommand{\arraystretch}{1.1}
\begin{tabular}{|c|c|c|c|c|c|}
\hline
GT$_p$ & ENS$_p$ & EDL$_p$ & GT$_\theta$ [$^\circ$] & ENS$_\theta$ [$^\circ$] & EDL$_\theta$ [$^\circ$] \\ \hline
0 & 0 & 0 & $[90,90]$ & $[-37.29,-14.31]$ & $[9.30,23.28]$ \\ \hline
1 & 1 & 1 & $[43,90]$ & $[42.36,46.40]$ & $[44.87,55.40]$ \\ \hline
2 & 2 & 2 & $[12,23]$ & $[11.50,24.32]$ & $[9.80,24.23]$ \\ \hline
\multicolumn{6}{|c|}{\textit{Closely spaced targets ($\Delta\theta < 5^\circ$)}} \\ \hline
2 & 2 & 2 & $[56,61]$ & $[55.47,67.68]$ & $[52.02,64.46]$ \\ \hline
2 & 2 & 2 & $[-27,-23]$ & $[-29.91,-17.87]$ & $[-29.45,-19.32]$ \\ \hline
\end{tabular}
\end{table}

\begin{table}[!t]
\centering
\caption{Representative ID examples: Corresponding uncertainty measures (ENS vs EDL) for the same samples in Table~\ref{tab:id_presence_doa}.}
\label{tab:id_uncertainty}
\footnotesize
\setlength{\tabcolsep}{3pt}
\renewcommand{\arraystretch}{1.1}
\begin{tabular}{|c|c|c|c|c|c|}
\hline
GT$_p$ & ENS$_p$ & EDL$_p$ & ENS$_\kappa$ & ENS$_\sigma$ [$^\circ$] & EDL$_\sigma$ [$^\circ$] \\ \hline
0 & 0 & 0 & $[0.66,2.27]$ & $[43.71,22.29]$ & $[47.91,47.35]$ \\ \hline
1 & 1 & 1 & $[38.81,2\mathrm{e}{-2}]$ & $[4.63,212.96]$ & $[12.54,14.06]$ \\ \hline
2 & 2 & 2 & $[56.27,57.20]$ & $[3.84,3.80]$ & $[11.33,11.05]$ \\ \hline
\multicolumn{6}{|c|}{\textit{Closely spaced targets ($\Delta\theta < 5^\circ$)}} \\ \hline
2 & 2 & 2 & $[24.11,19.26]$ & $[5.90,6.62]$ & $[13.84,13.81]$ \\ \hline
2 & 2 & 2 & $[20.14,20.10]$ & $[6.51,6.47]$ & $[14.89,17.09]$ \\ \hline
\multicolumn{6}{|p{0.9\linewidth}|}{
\footnotesize
$GT_p$, $ENS_p$, and $EDL_p$ denote the number of targets in the ROI cube.
$\theta$ and $\sigma$ denote the predicted DOA angles and the corresponding standard deviations (in degrees), respectively.
$\kappa$ denotes the aggregated concentration parameter obtained from the ENS.
} \\ \hline
\end{tabular}
\end{table}

The no-target (\(K=0\)) cases are consistently identified by both ENS and EDL, 
as seen in Table~\ref{tab:vm_edl_confusion} and Table~\ref{tab:id_presence_doa}, 
and primarily reflect the classification performance. 
Since uncertainty plays a more critical role in resolving ambiguities, 
the analysis focuses on single- and two-target scenarios, with emphasis 
on closely spaced two-target cases that present the primary ambiguity in DOA estimation.

The ID samples considered here include closely spaced targets with angular 
separations ranging from $5^\circ$ to $20^\circ$. 
Fig.~\ref{fig:ID_Uncertainty}(a) evaluates the DOA estimation accuracy of ENS and EDL within the ROI cube under in-distribution conditions. 
The accuracy is evaluated based on the absolute DOA error relative to the ground truth. 
For an angular error threshold of $\leq 5^\circ$, the ENS achieves accuracies of $\{80\%, 87\%\}$ for two-target and single-target cases, respectively, whereas EDL attains $\{74\%, 68\%\}$. 
As the error tolerance increases to $\leq 10^\circ$, the estimation accuracy exceeds $90\%$ for both methods across all target configurations.

Figs.~\ref{fig:ID_Uncertainty}(b)–(d) illustrate the corresponding total, 
aleatoric, and epistemic uncertainty statistics as detailed in Appendix~A. 
Across ID scenarios, ENS exhibits lower total and aleatoric uncertainty 
compared to EDL, while maintaining comparable or reduced epistemic uncertainty. 
These trends are consistent with the use of a von Mises likelihood, 
product-of-experts aggregation, and circular-statistics-based inference, 
as well as the incorporation of target presence information within the 
training objective. 

In contrast, EDL performs DOA estimation and uncertainty modeling via the 
Normal–Inverse–Gamma formulation, which does not explicitly enforce 
target-presence-dependent structure, and this may contribute to reduced 
accuracy in single-target cases relative to two-target scenarios. 
Additionally, the evidential formulation operates in a Euclidean output space, 
whereas the VM-based formulation provides a natural parametrization for 
directional variables. This difference can lead to broader uncertainty 
estimates in EDL compared to the more compact angular uncertainty 
captured by the ENS.

Overall, ENS provides more accurate and better-calibrated uncertainty 
estimates under in-distribution conditions, particularly for closely 
spaced multi-target scenarios.

The proposed models are not intended for super-resolution DOA estimation. 
In practice, targets exhibit spatial extent and multipath-induced scattering, 
leading to angular spreads on the order of $2^\circ$–$5^\circ$ even for 
single-target cases. For two-target scenarios with separations in the 
range $5^\circ$–$20^\circ$, each target retains a comparable intrinsic spread. 
In this regime, the challenge lies in robust DOA estimation and uncertainty 
quantification rather than resolution. ENS uncertainty estimates remain 
consistent with this expected angular spread, whereas EDL produces 
comparatively broader uncertainty intervals.

Since the single-target case benefits from explicit presence cues and 
typically exhibits lower ambiguity, uncertainty plays a comparatively 
less critical role than in multi-target scenarios. In contrast, closely 
spaced two-target cases require reliable uncertainty quantification to 
resolve overlapping angular responses. Accordingly, the out-of-distribution 
analysis focuses on closely spaced two-target scenarios, while the 
observations generalize to single-target cases.

\subsection{Out-of-Distribution (OOD) Analysis}
Consistent with the ID analysis, the OOD evaluation focuses on closely spaced two-target scenarios, which represent the most challenging cases for DOA estimation under distributional shift. 
The emphasis in the OOD setting is on jointly assessing estimation accuracy with respect to the ground truth and the associated predictive uncertainty.

To capture this trade-off, a risk-based metric is employed that combines the squared DOA estimation error and the corresponding uncertainty. 
Specifically, for a two-target scenario, the risk is defined as
\begin{equation}
\text{Risk} = \sqrt{\sum_{i=1}^{2}\left(\left|\theta_i - \mathrm{gt}_i\right|^{2} + \sigma_i^{2}\right)},
\label{eq:Risk_score}
\end{equation}
where $\theta_i$ and $\mathrm{gt}_i$ denote the estimated and ground-truth DOAs, respectively, and $\sigma_i$ represents the associated predictive standard deviation. 
This risk metric jointly penalizes large estimation errors and high predictive uncertainty, providing a unified measure for evaluating OOD robustness.

It is noted that lower risk values correspond to predictions with smaller estimation error and lower associated uncertainty, whereas higher risk values correspond to predictions with larger error, higher uncertainty, or both. 
With this risk metric defined, the subsequent section presents the risk–coverage characteristics under OOD conditions.

The OOD samples are constructed to induce controlled distributional shifts relative to the training data. 
A base OOD set uses targets at $80$\,m with an SNR of $-18$\,dB, which falls outside the training range yet remains relatively close to in-distribution conditions. 
To further introduce distributional mismatch, an additional perturbation is applied in the form of a deterministic phase shift corresponding to an excess propagation offset $\Delta r = 20$\,m. 
This choice introduces a nontrivial phase deviation relative to the trained range profiles without explicitly extending the target distance beyond the predefined scenario limits.

\begin{align}
\tilde{\mathbf{X}} &= \Big(\alpha\,\mathbf{X}\,e^{j\Delta\phi}\Big) * \mathbf{K} + \mathcal{N}, \label{eq:ood_perturb} \\
\Delta\phi &= \frac{4\pi \Delta r}{\lambda}, \quad \Delta r = 20\,\text{m}, \quad
\mathbf{K} =
\begin{bmatrix}
0.05 & 0.10 & 0.05 \\
0.10 & 0.40 & 0.10 \\
0.05 & 0.10 & 0.05
\end{bmatrix}. \nonumber
\end{align}

Here, $*$ denotes convolution applied channel-wise. 
The phase-perturbed signal is subjected to spatial leakage via the Gaussian kernel $\mathbf{K}$, followed by additive white Gaussian noise $\mathcal{N}$ such that the ROI cube attains an SNR of $-15$\,dB. 

These transformations introduce phase inconsistencies, spatial leakage, and elevated noise, thereby emulating effects analogous to multipath interference and range mismatch in a controlled manner.

\subsubsection{Risk-Coverage}
\begin{figure}[htbp]
    \centering
    \includegraphics[trim={4.7cm 0.5cm 5.5cm 0.5cm},width=0.45\linewidth]{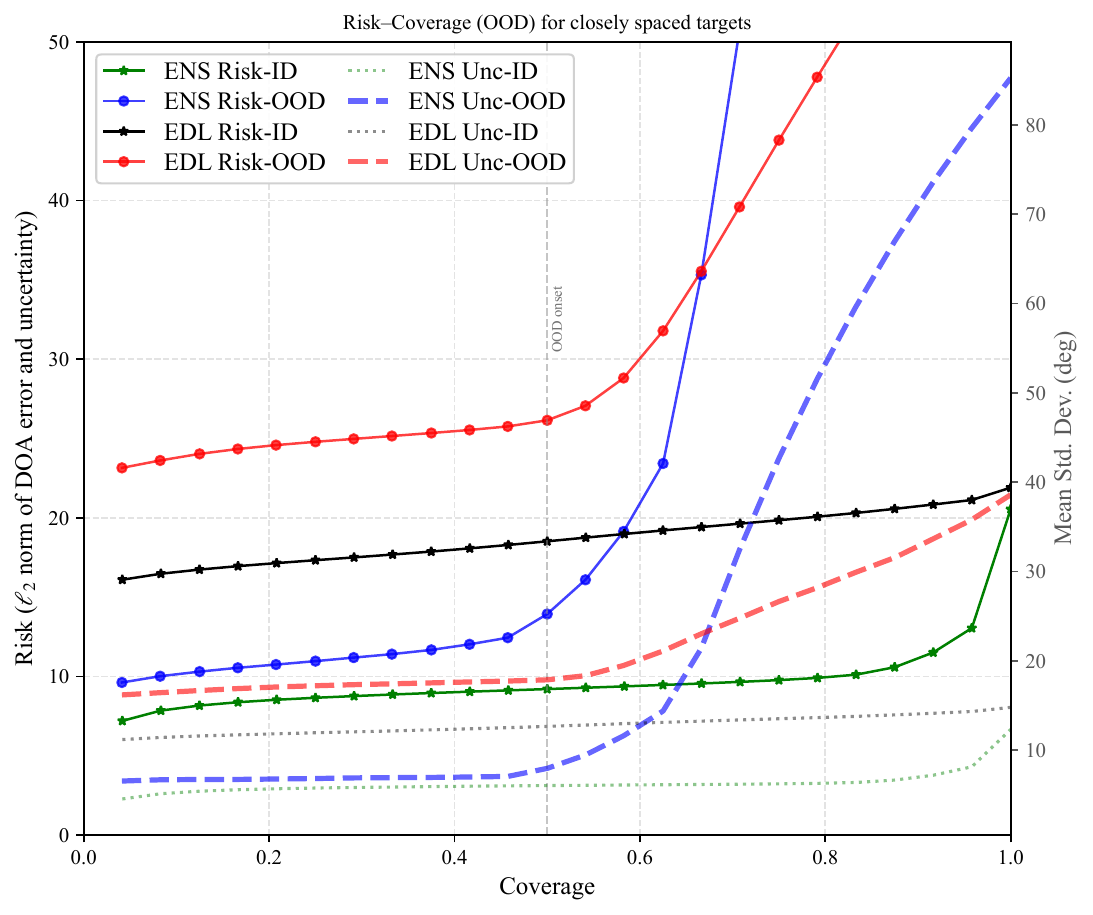}
    \caption{Risk coverage analysis for OOD samples at a range of $80$m under low‑SNR ($-18$dB) and under multipath‑faded conditions at the same range with SNR $-15$dB.}
    \vspace{-0.2cm}
    \label{fig:Risk_coverage}
\end{figure}

The risk–coverage characteristics are evaluated on the OOD samples generated as described in the preceding section (cf.~\eqref{eq:ood_perturb}). 
The risk–coverage curve shown in Fig.~\ref{fig:Risk_coverage} provides a joint perspective of prediction risk and coverage. 
The risk metric defined in~\eqref{eq:Risk_score} reflects the combined effect of estimation error and predictive uncertainty.

For the ENS model, the risk values corresponding to OOD samples that are closer to the in-distribution setting remain comparable to those observed under ID conditions (approximately $8^\circ$), while the OOD samples with stronger perturbations exhibit increased risk (approximately $10^\circ$) up to a coverage of about $45\%$, where a majority of the base OOD samples are concentrated. 
Beyond this coverage, a sharp increase in the risk metric is observed for the more challenging OOD samples affected by phase perturbation and leakage.

To provide additional context, the associated predictive standard deviation is also shown relative to ID behavior. 
Under ID conditions, the standard deviation remains within a narrow range of approximately $4$–$10^\circ$, whereas for OOD samples, the standard deviation increases significantly beyond $\sim50\%$ coverage, reaching values as high as $80^\circ$, indicating elevated predictive uncertainty under stronger distributional shifts.

In comparison, the EDL model exhibits higher baseline risk values under ID conditions (approximately $15^\circ$), with OOD samples further elevated (approximately $22^\circ$), indicating a separation between ID and OOD cases across the coverage range. 
However, beyond $\sim50\%$ coverage, the increase in risk is more gradual compared to ENS, with a corresponding steady increase in predictive uncertainty.

Overall, ENS demonstrates a wider dynamic range in both risk and uncertainty under OOD conditions, while EDL exhibits a more uniform separation between ID and OOD responses. 
This behavior highlights differing response characteristics of the two approaches under distributional shift.

The area under the risk–coverage curve (AURCC) is computed as the numerical integral of the risk metric with respect to coverage, following standard risk–coverage evaluation procedures~\cite{geifman2017selective}. Since the risk metric is expressed in degrees, the resulting AURCC values are also reported in degrees.
AURCC (ENS) ID = $9.2367$, AURCC (EDL) ID = $17.9310$; 
AURCC (ENS) OOD = $37.3238$, AURCC (EDL) OOD = $33.5205$.

It is worth noting that the observed AURCC trends are consistent with the differing shapes of the risk–coverage curves. 
For ENS, the risk values corresponding to ID and OOD samples remain close at lower coverage levels, followed by a sharp increase as coverage increases and more challenging OOD samples are included, leading to a higher cumulative contribution to AURCC.

In contrast, EDL exhibits a clearer separation between ID and OOD risk values at lower coverage levels, followed by a more gradual increase in risk. 
As a result, the cumulative risk (AURCC) grows more steadily. 
These observations indicate that while ENS captures a wider dynamic range in risk response under OOD conditions, EDL provides a more consistent separation between ID and OOD behavior across coverage levels.

\subsubsection{OOD detection using ROC/AUROC}

The additional complementary perturbations for OOD evaluation are introduced via progressive aperture reduction through missing receiver channels with injected noise. 
The base OOD sample at $80$\,m with an SNR of $-18$\,dB is used, which lies close to the in-distribution regime. 
Samples with ROI cube dimensions of $7 \times 17 \times 16$ are considered, where a subset of channels is replaced and additive white Gaussian noise is introduced to simulate OOD conditions for both ENS and EDL.

The evaluation is posed as a binary decision problem, where perturbed samples are labeled as $1$ (OOD) and the base samples as $0$ (ID-like). 
The Risk metric defined in~\eqref{eq:Risk_score} is used as the scoring function for each sample. 
Performance is summarized using the receiver operating characteristic (ROC) curve, obtained by plotting the true positive rate (TPR) against the false positive rate (FPR).

The area under the ROC curve (AUROC)~\cite{fawcett2006roc,hanley1982auc} provides a threshold-independent measure of separability. 
Importantly, AUROC depends only on the ranking induced by the Risk score and admits a probabilistic interpretation: it equals the probability that a randomly selected OOD sample is assigned a higher score than a randomly selected ID sample, with ties handled in the standard manner~\cite{hanley1982auc,bradley1997auc}.

\begin{figure}[htbp]
    \centering
    \includegraphics[trim={4.0cm 0.5cm 4.5cm 0.5cm}, width=0.45\linewidth]{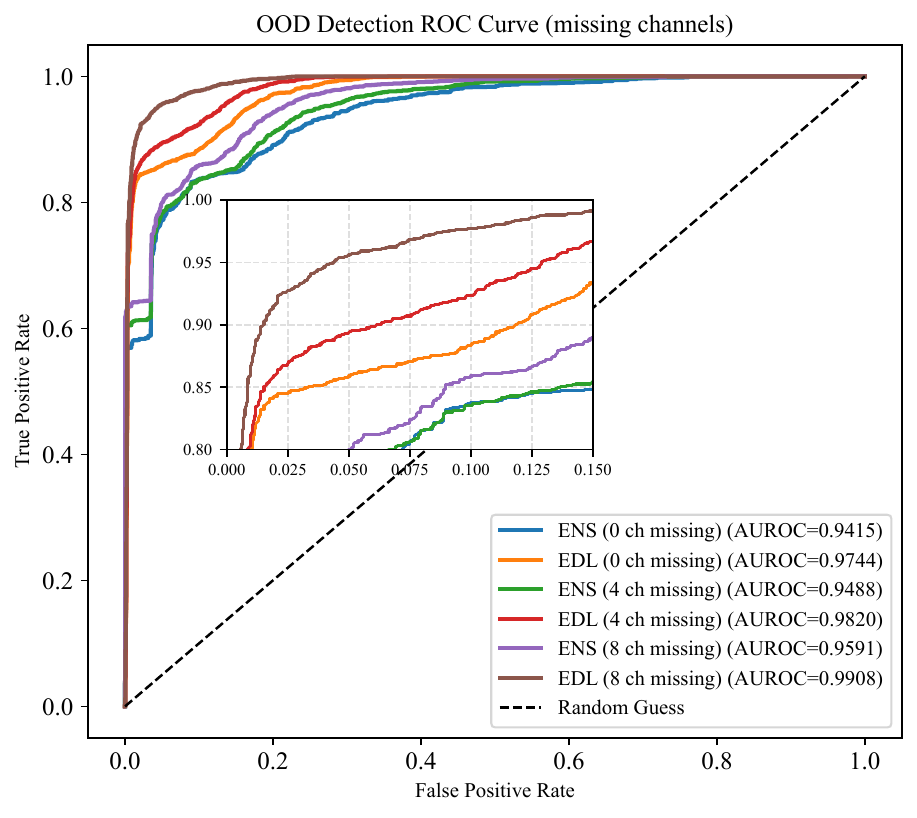}
    \caption{Sensor failure / partial array degradation OOD evaluation. ROC curves and AUROC for uncertainty‑based OOD detection with 0–8 missing virtual channels. In‑distribution data are generated at SNR $\in [-25,25]$ dB and 25 m, while OOD data correspond to SNR $-18$ dB at 80 m with additional multipath‑induced noise. The inset highlights the low‑FPR regime ($\leq 0.15$), where ID/OOD separability differences under aperture reduction are most pronounced.}
    \label{fig:AUROC_OOD}
\end{figure}

Fig.~\ref{fig:AUROC_OOD} compares ROC curves for OOD detection across multiple configurations. 
To evaluate progressively increasing OOD severity, no channel replacement is first applied to the base OOD samples, followed by configurations with $4$ and $8$ missing channels. 
The AUROC values reported in the legend summarize separability under each setting, with larger values indicating a more consistent ranking of OOD samples above ID samples. 
Consistent with the ROC definition, curves closer to the upper-left corner indicate higher TPR for a given FPR, whereas curves closer to the diagonal reflect weaker discrimination.

In this experiment, EDL achieves AUROC values ranging from $0.974$ to $0.991$, whereas ENS exhibits AUROC values between $0.941$ and $0.959$ under the same configurations. 
This behavior is consistent with differences in predictive representations between the two approaches. 
The VM-based ENS framework employs circular statistics to enforce angular consistency within a bounded domain (e.g., $-90^\circ$ to $90^\circ$), whereas the EDL framework models predictive uncertainty using a normal-inverse-gamma (NIG) formulation, resulting in a Student-$t$ predictive distribution that is not explicitly constrained in the circular domain.

These differences influence how uncertainty is expressed across perturbed inputs and, consequently, how samples are ranked under the Risk-based scoring function. 
Nevertheless, ENS exhibits comparable OOD detection performance in this aperture-reduction scenario.

For the array degradation experiments, ROC and AUROC are used to assess OOD detection capability, as the primary objective is to evaluate model confidence under corrupted inputs. 
For practical deployment, an operating threshold on the Risk score may be selected from the ROC curve corresponding to a desired false positive rate (FPR), enabling a trade-off between OOD detection sensitivity and false alarms. 
Both risk–coverage and ROC/AUROC analyses are presented for complementary OOD perturbations, including multipath-like distortions and aperture degradation. 
Across these settings, ENS demonstrates sharper variation in risk under distributional shift, while EDL exhibits more stable separation between ID and OOD samples. 
For deployment scenarios such as autonomous systems, this highlights a trade-off between sensitivity to high-uncertainty samples and consistency of OOD discrimination.

\section{Discussion and Practical Integration}

\subsection{Integration with Association Modules}

A key practical advantage of the ENS framework lies in its probabilistic representation of DOA using circular statistics. 
Each prediction is parameterized by a von Mises (VM) distribution with parameters $(\mu_{\mathrm{ens}}, \kappa_{\mathrm{ens}})$, directly representing the angular estimate and its associated concentration.

This representation enables seamless integration into downstream association modules. 
Given a track hypothesis with angular state $\theta_{\mathrm{tr}}$, the association likelihood can be evaluated analytically as
\begin{equation}
    p(\theta_{\mathrm{tr}} \mid \mu_{\mathrm{ens}}, \kappa_{\mathrm{ens}}) = \mathrm{VM}(\theta_{\mathrm{tr}}; \mu_{\mathrm{ens}}, \kappa_{\mathrm{ens}}),
    \label{eq:Association_VM}
\end{equation}
providing a closed-form directional likelihood in polar coordinates.

Furthermore, the intermediate sine–cosine representations $(S_j, C_j)$ used in circular statistics offer an alternative embedding that can be leveraged when association or tracking is performed in Cartesian or mixed coordinate systems, enabling consistent transformation without loss of angular structure.

In contrast, the EDL framework produces predictive outputs in Euclidean space through a Student-$t$ distribution derived from a normal-inverse-gamma formulation. 
While uncertainty estimates are available, this representation does not directly correspond to a circular probability density, requiring additional transformations or approximations for integration into angular association pipelines.

\subsection{Comparative Characteristics of ENS and EDL}

The ENS and EDL frameworks exhibit complementary strengths arising from their respective probabilistic representations and architectural design choices.

\subsection{Comparative Characteristics of ENS and EDL}

The ENS and EDL frameworks exhibit complementary strengths arising from their respective probabilistic representations and architectural choices.

\begin{table*}[t]
\centering
\caption{Comparison of ENS and EDL frameworks for DOA prediction, uncertainty modeling, and system integration.}
\label{tab:ens_edl_comparison}
\footnotesize
\setlength{\tabcolsep}{4pt}
\renewcommand{\arraystretch}{1.1}
\begin{tabular}{|c|p{6cm}|p{6cm}|}
\hline
\textbf{Aspect} & \textbf{ENS (VM-based Ensemble)} & \textbf{EDL (NIG-based)} \\ \hline

Geometric Consistency 
& Circular statistics; preserves wrap-around and bounded support ($[-90^\circ,90^\circ]$). 
& No explicit circular-domain modeling; operates in Euclidean space. \\ \hline

Uncertainty Modeling 
& $\kappa$ encodes confidence; epistemic uncertainty captured via ensemble diversity (3 VM models). 
& Aleatoric and epistemic uncertainty jointly modeled via NIG parameters. \\ \hline

Predictive Representation / Posterior Modeling 
& VM $(\mu_{\mathrm{ens}},\kappa_{\mathrm{ens}})$ on circular domain; approximate posterior via ensemble aggregation (VM-based) . 
& Normal-inverse-gamma (NIG) posterior; yields closed-form Student-$t$ predictive distribution in Euclidean space. \\ \hline

Association Integration 
& Direct likelihood evaluation in polar domain via \eqref{eq:Association_VM}. 
Also supports $(S_j,C_j)$ (\eqref{eq:Circular_agg_cart_polar}) for coordinate-consistent transformations. 
& No direct circular likelihood; requires mapping of Student-$t$ output to angular domain for association. \\ \hline

Behavior under ID 
& Lower predictive std.~dev. ($\sim 4^\circ$–$10^\circ$); tighter confidence bounds. 
& Higher baseline std.~dev. ($\sim 10^\circ$); requires calibration reference. \\ \hline

Behavior under OOD / Ranking (ROC/AUROC)
& Low baseline uncertainty under ID, but strong increase for severe perturbations; high sensitivity to difficult OOD samples, with less stable score ordering across thresholds.
& Higher baseline uncertainty under ID, requiring calibration relative to ID baseline; smoother variation leading to more stable ordering across thresholds and clearer ID–OOD separation, yielding slightly higher AUROC. \\ \hline

Computational Cost 
& $\sim$42K parameters per model; ensemble of 3 models increases memory and inference cost (parallelizable). 
& $\sim$42K parameters; single-model inference with lower memory and compute cost. \\ \hline

Deployment Suitability 
& Suitable for high-compute platforms (e.g., ADCU / centralized processing); benefits from GPU parallelization and multi-sensor likelihood fusion. 
& Suitable for resource-constrained embedded / edge devices due to compact architecture and lower compute requirements. \\ \hline

Practical Implication 
& Geometrically consistent probabilistic output enables direct integration into tracking/association pipelines. 
& Requires additional transformations for directional inference, increasing integration complexity. \\ \hline
Pipeline Role
& Enables joint detection–association integration via VM likelihood; supports fusion and consistent operation across coordinate systems, making it well-suited for end-to-end radar pipelines. 
& Primarily provides detection and uncertainty outputs; integration with tracking/association requires additional processing. \\ \hline

\end{tabular}
\end{table*}

The comparison in Table~\ref{tab:ens_edl_comparison} emphasizes the role of predictive representation in downstream radar processing. 
The ENS framework produces a von Mises distribution $(\mu_{\mathrm{ens}}, \kappa_{\mathrm{ens}})$ that enables direct probabilistic association through closed-form likelihood evaluation in the angular domain, allowing seamless integration with tracking modules.

In contrast, the EDL framework yields predictive uncertainty in Euclidean space via a Student-$t$ distribution derived from a normal-inverse-gamma formulation. 
While statistically grounded, it does not directly correspond to a circular-domain likelihood, requiring additional processing for integration with angular association pipelines. 
This distinction highlights a trade-off between statistical generality and geometric consistency in uncertainty-aware DOA estimation.
\section{Conclusion}

This work evaluates uncertainty-aware deep learning approaches for DOA estimation, comparing a von Mises–based ensemble (ENS) with an evidential deep learning (EDL) framework based on a normal-inverse-gamma formulation. Results across in-distribution and multiple out-of-distribution scenarios show that ENS achieves lower uncertainty under nominal conditions and exhibits strong sensitivity to severe perturbations, while EDL provides smoother uncertainty variation and slightly higher AUROC due to more stable ranking behavior. It is noted that the baseline uncertainty of EDL under in-distribution conditions is higher than that of ENS, and thus requires interpretation relative to this reference.

A key finding is that predictive representation directly impacts downstream integration. The ENS formulation produces a von Mises distribution $(\mu_{\mathrm{ens}}, \kappa_{\mathrm{ens}})$ enabling direct probabilistic association, whereas EDL outputs require additional transformations for angular inference. Overall, the results highlight a trade-off between geometric consistency and statistical generality, with ENS offering advantages for tightly coupled detection–tracking pipelines and EDL providing a compact and stable uncertainty formulation.
\begin{appendices}
\small{
\section{Circular-Statistics-Based Uncertainty Measures}
Let $\mu_j^{(e)}$ and $\kappa_j^{(e)}$ denote the predicted mean direction and concentration parameter of the $e$‑th ensemble member for the $j$‑th DOA, where $j \in \{1,2\}$ and $e \in \{1,2,3\}$.

The ensemble circular mean is computed as
\begin{align}
\bar{S}_j &= \frac{1}{3}\sum_{e=1}^{3} \sin\!\left(\mu_j^{(e)}\right), \quad \bar{C}_j = \frac{1}{3}\sum_{e=1}^{3} \cos\!\left(\mu_j^{(e)}\right),  \nonumber \\
\bar{\mu}_j &= \mathrm{atan2}\!\left(\bar{S}_j, \bar{C}_j\right).
\label{eq:Circular_agg_cart_polar}
\end{align}

The resultant length
\begin{equation}
\bar{R}_j = \sqrt{\bar{S}_j^2 + \bar{C}_j^2}
\end{equation}
is used to quantify ensemble dispersion. The corresponding circular variance and its Gaussian-equivalent dispersion are given by
\begin{equation}   
\mathrm{Var}_{\mathrm{epi},j} = 1 - \bar{R}_j, \quad \sigma_{\mathrm{epi},j} = \sqrt{-2\ln\!\left(\bar{R}_j\right)}  .
\end{equation}

For aleatoric uncertainty, the mean resultant length of each ensemble member is computed from the corresponding concentration parameter $\kappa_j^{(e)}$, modulated by the PoE weight $w_j^{(e)}$, given by
\begin{equation}
A_j^{(e)} = \frac{I_1\!\left(w_j^{(e)} \kappa_j^{(e)}\right)}{I_0\!\left(w_j^{(e)} \kappa_j^{(e)}\right)} .
\end{equation}
This leads to the ensemble-averaged circular variance and dispersion
\begin{align}
\mathrm{Var}_{\mathrm{ale},j} &= \frac{1}{3}\sum_{e=1}^{3} \left(1 - A_j^{(e)}\right), \nonumber \\
\sigma_{\mathrm{ale},j} &= \sqrt{-2\ln\!\left(1 - \mathrm{Var}_{\mathrm{ale},j}\right)} .
\end{align}

Assuming additive contributions in circular-variance space, the total uncertainty is computed as
\begin{align}
\mathrm{Var}_{\mathrm{tot},j} = \mathrm{Var}_{\mathrm{epi},j} + \mathrm{Var}_{\mathrm{ale},j}, \quad 
\sigma_{\mathrm{tot},j} = \sqrt{-2\ln\!\left(1 - \mathrm{Var}_{\mathrm{tot},j}\right)} .
\end{align}

The aggregated concentration parameter is then obtained as
\begin{equation}
\kappa_{\mathrm{tot},j} = A^{-1}\!\left(1 - \mathrm{Var}_{\mathrm{tot},j}\right),
\end{equation}
where $A^{-1}(\cdot)$ denotes the inverse of the mean resultant length of the von Mises distribution.
}
\end{appendices}

\begingroup\footnotesize 
    \bibliographystyle{IEEEtran} 
    \bibliography{references.bib}

\begin{thebibliography}{10}
\providecommand{\url}[1]{#1}
\csname url@samestyle\endcsname
\providecommand{\newblock}{\relax}
\providecommand{\bibinfo}[2]{#2}
\providecommand{\BIBentrySTDinterwordspacing}{\spaceskip=0pt\relax}
\providecommand{\BIBentryALTinterwordstretchfactor}{4}
\providecommand{\BIBentryALTinterwordspacing}{\spaceskip=\fontdimen2\font plus
\BIBentryALTinterwordstretchfactor\fontdimen3\font minus
  \fontdimen4\font\relax}
\providecommand{\BIBforeignlanguage}[2]{{%
\expandafter\ifx\csname l@#1\endcsname\relax
\typeout{** WARNING: IEEEtran.bst: No hyphenation pattern has been}%
\typeout{** loaded for the language `#1'. Using the pattern for}%
\typeout{** the default language instead.}%
\else
\language=\csname l@#1\endcsname
\fi
#2}}
\providecommand{\BIBdecl}{\relax}
\BIBdecl

\bibitem{kay1993fundamentals}
S.~M. Kay, \emph{Fundamentals of statistical signal processing: detection
  theory}.\hskip 1em plus 0.5em minus 0.4em\relax Prentice-Hall, Inc., 1993.

\bibitem{richards2014fundamentals}
M.~A. Richards, \emph{Fundamentals of radar signal processing}.\hskip 1em plus
  0.5em minus 0.4em\relax McGraw-Hill Education, 2014.

\bibitem{sun2020mimo}
S.~Sun, A.~P. Petropulu, and H.~V. Poor, ``{MIMO} radar for advanced
  driver-assistance systems and autonomous driving: Advantages and
  challenges,'' \emph{IEEE Signal Process. Magazine}, vol.~37, no.~4, pp.
  98--117, 2020.

\bibitem{skolnik1980introduction}
M.~I. Skolnik, ``Introduction to radar systems,'' \emph{New York}, 1980.

\bibitem{wang2016spatial}
H.-N. Wang, Y.-W. Huang, and S.-J. Chung, ``Spatial diversity 24-{GHz FMCW}
  radar with ground effect compensation for automotive applications,''
  \emph{IEEE Trans. Veh. Techn.}, vol. 66(2), pp. 965--973, 2016.

\bibitem{barton1997radar}
D.~K. Barton and S.~Leonov, ``Radar technology encyclopedia, artech house,''
  \emph{Inc., Ed., Norwood, MA (USA)}, 1997.

\bibitem{series2014systems}
M.~Series, ``Systems characteristics of automotive radars operating in the
  frequency band 76--81 ghz for intelligent transport. systems applications,''
  \emph{Recommendation ITU-R, M}, pp. 2057--1, 2014.

\bibitem{Qian2025SensorFusionSurvey}
H.~Qian, M.~Wang, M.~Zhu, and H.~Wang, ``A review of multi-sensor fusion in
  autonomous driving,'' \emph{Sensors}, vol.~25, no.~19, p. 6033, 2025.

\bibitem{Dowling2026CentralizedRadar}
L.~Dowling, N.~Shigihalli, S.~Murray, and B.~Fathi, ``How centralized radar
  processing enables safer, smarter autonomous driving,'' \emph{NVIDIA
  Developer Technical Blog}, 2026.

\bibitem{Gawlikowski2023Survey}
J.~Gawlikowski, F.~Kahl, M.~Heine, and et~al., ``A survey of uncertainty
  quantification in deep learning,'' \emph{IEEE Transactions on Neural Networks
  and Learning Systems}, 2023.

\bibitem{Kazazi2025RDCFAR}
J.~Kazazi, M.~Kamarei, and M.~Fakharzadeh, ``{RD-CFAR}: Fast and accurate
  constant false alarm rate algorithm for automotive radar applications,''
  \emph{IEEE Signal Processing Letters}, 2025.

\bibitem{CFARComparison2025IEEEAccess}
A.~Unknown, ``Comparison of cfar algorithms in real-time target detection for
  automotive radar,'' \emph{IEEE Access}, 2025.

\bibitem{Wang2023AdaptiveCFAR}
Y.~Wang, H.~Liu, and X.~Li, ``Adaptive two-dimensional cfar detection for
  automotive radar in nonhomogeneous environments,'' \emph{IEEE Transactions on
  Vehicular Technology}, 2023.

\bibitem{Rohling2024KACFAR}
H.~Rohling and M.~M{\"u}nster, ``Knowledge-aided cfar detection for automotive
  radar applications,'' \emph{IEEE Transactions on Aerospace and Electronic
  Systems}, 2024.

\bibitem{BarShalom2009Tracking}
Y.~Bar-Shalom, X.~R. Li, and T.~Kirubarajan, \emph{Estimation with Applications
  to Tracking and Navigation}.\hskip 1em plus 0.5em minus 0.4em\relax Wiley,
  2001.

\bibitem{Mahler2007PHD}
R.~Mahler, \emph{Statistical Multisource-Multitarget Information Fusion}.\hskip
  1em plus 0.5em minus 0.4em\relax Artech House, 2007.

\bibitem{Yu2025CloselySpacedAssociation}
Z.~Yu, Z.~Jin, T.~Sun, J.~Ding, J.~Li, and Q.~Guo, ``Closely spaced
  multi-target association and localization using br and aoa measurements in
  distributed {MIMO} radar systems,'' \emph{Remote Sensing}, vol.~17, no.~6,
  2025.

\bibitem{Delamou2023DLRadar}
M.~Delamou, A.~Bazzi, and M.~Chafii, ``Deep learning-based estimation for
  multitarget radar detection,'' \emph{arXiv preprint arXiv:2305.05621}, 2023.

\bibitem{Roldan2024SeeFurtherThanCFAR}
I.~R. Montero, A.~Palffy, J.~F.~P. Kooij, D.~M. Gavrila, F.~Fioranelli, and
  A.~Yarovoy, ``See further than cfar: A data-driven radar detector trained by
  lidar,'' in \emph{Proceedings of the IEEE Radar Conference (RadarConf)},
  2024.

\bibitem{Bauw2026ComplexNNRadar}
M.~Bauw, ``Detecting radar targets in range profiles with partially
  complex-valued neural networks,'' \emph{arXiv preprint arXiv:2602.09597},
  2026.

\bibitem{Kaiser2021CVNNRadar}
K.~Kaiser, J.~Daugalas, J.~L{\'o}pez-Randulfe, A.~Knoll, R.~Weigel, and
  F.~Lurz, ``Complex-valued neural networks for millimeter-wave fmcw radar
  angle estimation,'' in \emph{European Radar Conference}, 2021.

\bibitem{kulkarni2025kan}
V.~Kulkarni, V.~Reddy, and N.~Maheshwari, ``Kan-powered large-target detection
  for automotive radar,'' \emph{IEEE Transactions on Radar Systems}, 2025.

\bibitem{kulkarni2025gamma}
V.~Kulkarni and V.~Reddy, ``Gamma-based statistical modeling for extended
  target detection in mmwave automotive radar,'' \emph{arXiv preprint
  arXiv:2509.26573}, 2025.

\bibitem{wei2022area}
Z.~Wei, B.~Li, T.~Feng, Y.~Tao, and C.~Zhao, ``Area-based {CFAR} target
  detection for automotive millimeter-wave radar,'' \emph{IEEE Trans. Vehicular
  Technology}, vol.~72, no.~3, pp. 2891--2906, 2022.

\bibitem{kulkarni2023detection}
V.~Kulkarni, V.~Reddy, and A.~Dixit, ``Detection of close-proximity automotive
  targets using lstm,'' in \emph{IEEE Radar Conf.}, 2023, pp. 1--6.

\bibitem{Liang2025NoiseIntegralSBL}
Y.~Liang, X.~Zheng, W.~Meng, and J.~Li, ``Noise integral-based sparse bayesian
  learning for doa estimation using grid pruning and adaptation,'' \emph{IET
  Electronics Letters}, 2025.

\bibitem{Li2024BayesianUnfoldedDOA}
N.~Li, X.~Zhang, F.~Lv, B.~Zong, and W.~Feng, ``A bayesian deep unfolded
  network for off-grid direction-of-arrival estimation,'' \emph{Electronics},
  vol.~13, no.~11, 2024.

\bibitem{Chen2020SBLDOA}
S.~Liu, L.~Tang, Y.~Bai, and X.~Zhang, ``A sparse bayesian learning-based doa
  estimation method with the kalman filter in mimo radar,'' \emph{Electronics},
  vol.~9, no.~2, p. 347, 2020.

\bibitem{Sensoy2018EDL}
M.~Sensoy, L.~Kaplan, and M.~Kandemir, ``Evidential deep learning to quantify
  classification uncertainty,'' \emph{Advances in Neural Information Processing
  Systems (NeurIPS)}, 2018.

\bibitem{Amini2020EvidentialRegression}
A.~Amini, W.~Schwarting, A.~Soleimany, and D.~Rus, ``Deep evidential
  regression,'' \emph{Advances in Neural Information Processing Systems
  (NeurIPS)}, 2020.

\bibitem{Gal2016MCDropout}
Y.~Gal and Z.~Ghahramani, ``Dropout as a bayesian approximation: Representing
  model uncertainty in deep learning,'' in \emph{Proceedings of the
  International Conference on Machine Learning (ICML)}, 2016.

\bibitem{lakshminarayanan2017simple}
B.~Lakshminarayanan, A.~Pritzel, and C.~Blundell, ``Simple and scalable
  predictive uncertainty estimation using deep ensembles,'' \emph{Advances in
  neural information processing systems}, vol.~30, 2017.

\bibitem{wilson2020bayesian}
A.~G. Wilson and P.~Izmailov, ``Bayesian deep learning and a probabilistic
  perspective of generalization,'' \emph{Advances in neural information
  processing systems}, vol.~33, pp. 4697--4708, 2020.

\bibitem{mardia2000directional}
K.~V. Mardia and P.~E. Jupp, \emph{Directional Statistics}.\hskip 1em plus
  0.5em minus 0.4em\relax Wiley, 2000.

\bibitem{hinton2002training}
G.~E. Hinton, ``Training products of experts by minimizing contrastive
  divergence,'' \emph{Neural computation}, vol.~14, no.~8, pp. 1771--1800,
  2002.

\bibitem{woo2018cbam}
S.~Woo, J.~Park, J.-Y. Lee, and I.~S. Kweon, ``Cbam: Convolutional block
  attention module,'' in \emph{Proceedings of the European Conference on
  Computer Vision (ECCV)}, 2018, pp. 3--19.

\bibitem{kuhn1955hungarian}
H.~W. Kuhn, ``The hungarian method for the assignment problem,'' \emph{Naval
  research logistics quarterly}, vol.~2, no. 1-2, pp. 83--97, 1955.

\bibitem{amini_edl_github}
A.~Amini and Contributors, ``Deep evidential regression (nig) implementation,''
  \url{https://github.com/aamini/evidential-deep-learning}.

\bibitem{mardia_jupp}
K.~V. Mardia and P.~E. Jupp, \emph{Directional Statistics}.\hskip 1em plus
  0.5em minus 0.4em\relax John Wiley \& Sons, 2000.

\bibitem{geifman2017selective}
Y.~Geifman and R.~El-Yaniv, ``Selective classification for deep neural
  networks,'' in \emph{NeurIPS}, 2017.

\bibitem{fawcett2006roc}
T.~Fawcett, ``An introduction to roc analysis,'' \emph{Pattern Recognition
  Letters}, vol.~27, no.~8, pp. 861--874, 2006.

\bibitem{hanley1982auc}
J.~A. Hanley and B.~J. McNeil, ``The meaning and use of the area under a
  receiver operating characteristic (roc) curve,'' \emph{Radiology}, vol. 143,
  no.~1, pp. 29--36, 1982.

\bibitem{bradley1997auc}
A.~P. Bradley, ``The use of the area under the roc curve in the evaluation of
  machine learning algorithms,'' \emph{Pattern Recognition}, vol.~30, no.~7,
  pp. 1145--1159, 1997.

\end{thebibliography}
\endgroup

\end{document}